\def\masy{~\textrm{mas~yr}^{-1}}

\def\eqi{\begin{equation}}
\def\eqf{\end{equation}}
\def\eqia{\begin{eqnarray}}
\def\eqfa{\end{eqnarray}}

\def\lb#1{\label{#1}}

\def\ton#1{\left(#1\right)}

\documentclass[onecolumn]{aastex}

\usepackage[title]{appendix}
\usepackage[table,xcdraw]{xcolor}
\usepackage{multirow}
\usepackage{rotating,tabularx}
\usepackage{float}
\usepackage{enumerate}
\usepackage[polutonikogreek,english]{babel}
\usepackage{amsmath,textgreek,w-greek,wasysym}
\usepackage{amsthm}
\usepackage{amscd,lineno}
\usepackage{amssymb,dsfont}
\usepackage{graphicx,epsfig}
\usepackage{txfonts,tablefootnote}
\usepackage{ragged2e}
\usepackage{xr-hyper}

\RequirePackage{color}

\newcommand{\emaila}{lorenzo.iorio@libero.it}

\allowdisplaybreaks[1]

\begin{document}

\title{Calculation of the Uncertainties in the Planetary Precessions with the Recent EPM2017 Ephemerides and their Use in Fundamental Physics and Beyond}

\shortauthors{L. Iorio}

\author{Lorenzo Iorio\altaffilmark{1} }
\affil{Ministero dell'Istruzione, dell'Universit\`{a} e della Ricerca
(M.I.U.R.)-Istruzione
\\ Permanent address for correspondence: Viale Unit\`{a} di Italia 68, 70125, Bari (BA),
Italy}

\email{\emaila}

\begin{abstract}
I  tentatively compile the formal uncertainties in the secular rates of change of the orbital elements $a,~e,~I,~\Omega$ and $\varpi$ of the planets of the solar system from the recently released formal errors in $a$ and the nonsingular elements $h,~k,~p$ and $q$ estimated for the same bodies with the EPM2017 ephemerides by E.\,V. Pitjeva and N.\,P. Pitjev. The highest accuracies occur for the inner planets and Saturn in view of the extensive use of radiotechnical data collected over the last decades. For the inclination $I$, node $\Omega$ and perihelion $\varpi$ of Mercury and Mars, I obtain accuracies  $\upsigma_{\dot I},~\upsigma_{\dot\Omega},~\upsigma_{\dot \varpi}\simeq 1-10~\upmu\textrm{as~cty}^{-1}$, while for Saturn they are  $\upsigma_{\dot I},~\upsigma_{\dot\Omega},~\upsigma_{\dot \varpi}\simeq 10~\upmu\textrm{as~cty}^{-1}-1~\textrm{mas}~\textrm{cty}^{-1}$.  As far as the semimajor axis $a$ is concerned, its rates for the inner planets are accurate to the $\upsigma_{\dot a}\simeq 1-100~\textrm{mm~cty}^{-1}$ level, while for Saturn I obtain  $\upsigma_{\dot a}\simeq 17~\textrm{m~cty}^{-1}$. In terms of the parameterized post-Newtonian (PPN) parameters $\beta$ and $\gamma$, a formal error as little as $8~\upmu\textrm{as~cty}^{-1}$ for the Hermean apsidal rate corresponds to a $\simeq 2\times 10^{-7}$ bias in the combination $\ton{2 + 2\gamma- \beta}/3$ parameterizing the Schwarzschild-type periehlion precession of Mercury. The realistic uncertainties of the planetary precessions may be up to one order of magnitude larger. I  discuss their potential multiple uses in fundamental physics, astronomy, and planetology.
\end{abstract}

keywords{
gravitation $-$ celestial mechanics $-$ astrometry $-$ ephemerides
}

%

\section{Introduction}\lb{intro}
Knowing accurately the orbital motions of the major bodies of our solar system \citep{2007SoSyR..41..265K} has historically played a fundamental role over the years in putting to the test alternative theories of gravity to those that had been deemed as established from time to time. Just think of the anomalous perihelion precession of Mercury at that time \citep{LeVer1859} and its successful explanation by \citet{1915SPAW...47..831E} with his newly born general theory of relativity (GTR); for recent overviews of its status and perspectives after one century from its publication, see, e.g., \citet{2016Univ....2...23D} and references therein. Moreover, the most accurate tests of GTR have been performed so far just in the solar system arena, although it can probe its weak-field and slow-motion approximation. Indeed, binary pulsar systems, representing the most direct competitors   in terms of the obtainable accuracy of the GTR validity, currently allow to reach the $5\times 10^{-4}$ level \citep{2016IJMPD..2530029K}, which is about one order of magnitude less accurate than the most recent solar system-based results \citep{2014CeMDA.119..237P,2015CeMDA.123..325F,2017AJ....153..121P,2018NatureG}. On the other hand, as far as GTR itself is concerned, not all of its features of the motion of order $\mathcal{O}\ton{c^{-2}}$, where $c$ is the speed of light in vacuum, have yet been  tested with planetary motions. Indeed, a novel GTR-induced $N-$body effect was recently predicted by \citet{2018PhRvL.120s1101W}, while the gravitomagnetic Lense$-$Thirring orbital precessions due to the Sun's angular momentum, $\mathbf{S}$ \citep{2011Ap&SS.331..351I}, has escaped from detection so far due to its minuteness; both of them may become measurable in the next few years by means of the Hermean orbital precessions when the data collected by the ongoing spacecraft-based \textit{BepiColombo} mission to Mercury will be finally collected and analyzed. Furthermore, there are currently several modified models of gravity, put forth mainly to cope with issues arisen at galactic (dark matter) and cosmological (dark energy) scales, which, as fortunate by-products, predict effects on the orbital motion of a test particle which could be measured or, at least, constrained in our solar system; see, e.g., \citet{2003PhRvD..67f4002L}, \citet{2009MNRAS.399..474M},
\citet{2011MNRAS.412.2530B},
\citet{2012PhRvD..86d4002G},
\citet{2012CQGra..29w5027H},
\citet{2012MNRAS.426..673M},
\citet{2014PhRvD..89j2002H},
\citet{2017PhLB..769..281L}, and
\citet{2018CQGra..35qLT01W}.
Last but not least, the location of the recently hypothesized new, remote planet of the solar system \citep{2016AJ....151...22B,2016ApJ...824L..23B}, provisionally known as Planet Nine or Telisto, can be effectively constrained by means of the orbital precessions of the other known major bodies of the solar system \citep{2016A&A...587L...8F,2017Ap&SS.362...11I}.

 During the last 15 years or so, two independent teams of astronomers, led by E.\,V. Pitjeva and A. Fienga, engaged in the production of more and more accurate planetary ephemerides (Ephemeris of Planets and the Moon (EPM) and Int\'{e}grateur Num\'{e}rique Plan\'{e}taire de l'Observatoire de Paris (INPOP), respectively), have determined increasingly accurate supplementary perihelion\footnote{\citet{2011CeMDA.111..363F} also  released supplementary precessions $\Delta\dot\Omega$ of the nodes determined with the INPOP10a ephemerides.} precessions, $\Delta\dot\varpi$, of all of the planets of the solar system by processing increasingly extensive and precise data records of all types. Such supplementary rates are usually estimated by confronting, mainly in a least-square sense, a suite of models, accurate to the first post-Newtonian order $\mathcal{O}\ton{c^{-2}}$, describing the dynamics\footnote{Until the recent advent of the EPM2017 ephemerides \citep{2018AstL...44..554P}, the post-Newtonian gravitomagnetic field of the Sun had never been modeled, with the exception of \citet{2017AJ....153..121P}}
 of the major and of most of the minor bodies of the solar system like the asteroids and the trans-Neptunian Objects (TNOs), the propagation of the electromagnetic waves and the functioning of the measurement devices (spacecraft transponders, etc.) with long data records covering about the last century or so. Thus, in principle, $\Delta\dot\varpi$ accounts for any unmodeled and mismodeled features of motion induced, e.g., by some putative exotic interaction of gravitational origin. However, the signatures of the latter ones may have been somewhat removed in the data reduction procedure, having been partly absorbed in the estimated values of, say, the initial state vectors \citep{2012CQGra..29w5027H}. Thus, in some cases which, however, cannot be established a priori, the bounds inferred by a straightforward comparison of $\Delta\dot\varpi$ to the corresponding theoretical perihelion precessions $\dot\varpi_\textrm{th}$ predicted by the dynamical models from time to time of interest may, perhaps, be too optimistically tight, at least to a certain extent. As  such, dedicated (and time-consuming) analyses performed by reprocessing the same data records by explicitly modeling the dynamical features under consideration should be performed, and the correlations among the estimated parameters in the resulting covariance matrix should be inspected.
 It is a task that would be unrealistic to think that it can be implemented every time one wants to test this or that model, also because it requires specific skills that, basically, only the astronomers responsible for the generation of the planetary ephemerides, whose priorities are often different, have.
 Be that as it may, the corrections $\Delta\dot\varpi$ to the standard perihelion rates, which, so far, have always been statistically compatible with zero, are long used  by researchers all over the world to put constraints on a variety of modified models of gravity and other dynamical features of motion just by straightforwardly comparing them to theoretical precessions; to fully realize the extent of such a practice, just consult the citation records of, say, \citet{2011CeMDA.111..363F} and \citet{2013MNRAS.432.3431P} in the SAO/NASA ADS database.
\section{The Uncertainties in the Planetary Orbital Rates of Change from the EPM2017 Ephemerides}
Recently, \citet{2018AstL...44..554P} released the EPM2017 ephemerides\footnote{See also http://iaaras.ru/en/dept/ephemeris/epm/2017/ on the Internet.} which, among other things, also include the Lense$-$Thirring field of the Sun in their dynamical models and rely upon the data collected by the spacecraft MESSENGER at Mercury (2011-2015). For some reason, \citet{2018AstL...44..554P} did not provide updated values of the supplementary perihelion advances $\Delta\dot\varpi$, limiting themselves to yield the statistical, formal uncertainties in the estimated values of the semimajor axes $a$ and of the nonsingular orbital elements  $h,~k,~p$ and $q$ of all the planets along with Pluto in their Table~3. In view of the previously outlined importance, in Table~\ref{tavola1} I tentatively compiled the formal uncertainties in the long-term rates of change of the Keplerian orbital elements $a,~e,~I,~\Omega$ and $\varpi$ of the eight planets of the solar system and of Pluto as follows.
\begin{table}[!htb]
\begin{center}
\caption{Formal Uncertainties $\upsigma_{\dot a},~\upsigma_{\dot e},~\upsigma_{\dot I},~\upsigma_{\dot\Omega},~\upsigma_{\dot\varpi}$ in the Secular Rates of Change of the Semimajor Axis $a$, Eccentricity $e$, Inclination $I$, Longitude of the Ascending Node $\Omega$, and Longitude of Perihelion $\varpi$ of the Planets of Our Solar System  Tentatively Computed from the Formal Errors in the Nonsingular Orbital Elements Listed in Table~3 of \citet{2018AstL...44..554P} and the Temporal Lengths of the Data Records for Each Planet Listed in Table~2 and Figure~2 of Pitjeva \& Pitjev\,(2018; See the Text for Details).
}\lb{tavola1}
\begin{tabular}{cccccccccc}
  \hline
 & Mercury & Venus  & Earth & Mars & Jupiter & Saturn & Uranus & Neptune & Pluto \\
\hline
$\upsigma_{\dot a}$      & $0.003$  & $0.092$  & $0.062$  & $0.099$   & $4650$   & $16.936$ & $31630.3$ & $288035$  & $790006$\\
$\upsigma_{\dot e}$      & $0.0006$ & $0.0021$ & $0.0028$ & $0.0002$  & $2.016$   & $0.0023$ & $2.732$   & $10.696$  & $15.201$\\
$\upsigma_{\dot I}$      & $0.003$  & $0.050$  & $-$      & $0.002$   & $20.41$   & $0.063$  & $3.827$   & $4.733$   & $3.601$\\
$\upsigma_{\dot\Omega}$  & $0.024$  & $0.862$  & $-$      & $0.055$   & $959.1$   & $1.806$  & $269.177$ & $147.214$ & $4.917$\\
$\upsigma_{\dot\varpi}$  & $0.008$  & $0.315$  & $0.033$  & $0.003$   & $33.9$   & $0.067$  & $47.998$  & $1289.26$ & $60.810$\\
\hline
\end{tabular}
\begin{tabular}{>{\RaggedRight}p{\linewidth}}
\textbf{Note.} For the Earth, a spacecraft-based data record 21 yr long was assumed (see the text).   The actual uncertainties may be up to one order of magnitude larger. The units are metres  per  century $\ton{\textrm{m}\,\textrm{cty}^{-1}}$ for $\upsigma_{\dot a}$ and milliarcseconds per century $\ton{\textrm{mas}\,\textrm{cty}^{-1}}$ for $\upsigma_{\dot e}$,\,$\upsigma_{\dot I}$,\,$\upsigma_{\dot\Omega}$,\,$\upsigma_{\dot \varpi}$.
The mean ecliptic and equinox at J2000.0 were used for the computation of $\upsigma_{\dot e}$,\,$\upsigma_{\dot I}$,\,$\upsigma_{\dot\Omega}$,\,$\upsigma_{\dot\varpi}$.
\end{tabular}
\end{center}
\end{table}
First, analytical expressions for $e,~I,~\Omega$ and $\varpi$ as functions of the nonsingular elements $h = e\sin \varpi,~k = e\cos\varpi,~p =\sin I\sin\Omega$ and $q = \sin I\cos\Omega$ were calculated. Then, they were  differentiated with respect to $h,~k,~p$ and $q$ in order to calculate the errors in the root-sum-square fashion as, say, $\upsigma_I = \sqrt{ \ton{\partial I/\partial p}^2 \upsigma_p^2 + \ton{\partial I/\partial q}^2 \upsigma_q^2 }$, etc., where
$\upsigma_h,~\upsigma_k,~\upsigma_p,~\upsigma_q$ are the formal errors quoted in Table~3 of \citet{2018AstL...44..554P}.
Finally, the ratios of the previously computed errors $\upsigma_I,~\upsigma_e,~\upsigma_\Omega,~\upsigma_\varpi$ and of $a$ as per Table~3 of \citet{2018AstL...44..554P} to the lengths $\Delta t$ of the data records listed in Table~2 of \citet{2018AstL...44..554P} were taken for each planet, with some exceptions explained below for Venus and Jupiter for which no spacecraft-based data records spanning decades are available.
As pointed out by \citet{2018AstL...44..554P} themselves, the actual uncertainties may be up to one order of magnitude larger.
As far as the Euler-type angles $I,~\Omega$ and $\varpi$ determining the  orientation of the orbit in space are concerned, the inclination $I$ exhibits the most accurate precessions whose uncertainties may be as little as $\simeq \upmu\textrm{as~cty}^{-1}$ for Mercury and Mars, while for Saturn it is at the $\simeq 10~\upmu\textrm{as~cty}^{-1}$ level. The perihelion precessions are, essentially, at the same level of accuracy. The uncertainties in the rates of change of the nodes of Mercury and Mars are $\simeq 10~\upmu\textrm{as~cty}^{-1}$, while for Saturn it is of the order of $\simeq 1~\textrm{mas~cty}^{-1}$.

The present approach was tested with the available information from the EPM2011 ephemerides about the perihelia of all the planets apart from Uranus, Neptune, and Pluto. Indeed, Table~3 of \citet{2018AstL...44..554P} displays the formal errors of $a$ and the nonsingular elements also for such earlier ephemerides; Table~4 of \citet{2013MNRAS.432.3431P} explicitly releases the EPM2011-based supplementary perihelion precessions $\Delta\dot\varpi$ along with their uncertainties (in mas yr$^{-1}$), while the data intervals used are quoted in Table~1 and Table~3 of \citet{2013MNRAS.432.3431P}. Thus, it is possible to apply our approach to the EPM2011 uncertainties of Table~3 of \citet{2018AstL...44..554P} with the temporal intervals of Table~1 and Table~3 of \citet{2013MNRAS.432.3431P} in order to calculate our own uncertainties, $\upsigma_{\dot\varpi}$, in the perihelion precessions and compare them with those in Table~4 of \citet{2013MNRAS.432.3431P}.
The resulting agreement is good as long as the temporal intervals, $\Delta t$, with which the rates of change are to be constructed are chosen wisely. For Venus, our strategy is able to reproduce the uncertainty $\upsigma_{\dot\varpi}$ listed in Table~4 of \citet{2013MNRAS.432.3431P} provided that the Magellan or the Venus Express (VEX) data intervals covering $\Delta t = 3-4~\textrm{yr}$ reported in Table~1 of \citet{2013MNRAS.432.3431P} are adopted. Thus, I followed the same approach with the EPM2017 by dividing the computed uncertainty in the orbital elements of Venus by the VEX temporal interval of Figure~2 of \citet{2018AstL...44..554P} which is 7 yr long.
As far as the Earth is concerned, it turns out that, in order to obtain the same uncertainty $\upsigma_{\dot\varpi}$ published in \citet{2013MNRAS.432.3431P}, a time span of $\Delta t = 15~\textrm{yr}$ should be adopted. Thus, when using the uncertainties for EPM2017 in order to compile Table~\ref{tavola1}, a data record of $\Delta t = 21~\textrm{yr}$ was used. In case of Jupiter, I am able to obtain the error  $\upsigma_{\dot\varpi}$ quoted in Table~4 of \citet{2013MNRAS.432.3431P} if the time span of $\Delta t = 8~\textrm{yr}$ of the Jovian orbiter Galileo is assumed. Since \citet{2018AstL...44..554P}  did not use the most recent data from Juno, also in obtaining Table~\ref{tavola1} I also used the Galileo data interval.

In order to better place in context the figures of Table~\ref{tavola1}, let us remark that an accuracy as good as $\upsigma_{\dot\varpi}=8~\upmu\textrm{as~cty}^{-1}$ for the perihelion precession of Mercury, which is better than the value quoted in Table ~4 of \citet{2013MNRAS.432.3431P} by a factor of about 300, corresponds to an uncertainty as little as $2\times 10^{-7}$ in the combination $\ton{2  + 2\gamma - \beta}/3$ of the PPN parameters $\gamma$ and $\beta$ multiplying the time-honored Schwarzschild-type Hermean precession of $42.98~\textrm{arcsec~cty}^{-1}$.
After all, an inspection of Table~3 of \citet{2018AstL...44..554P} shows that an improvement of more than two orders of magnitude occurred for the nonsingular orbital elements $h,~k$ of Mercury in the transition from the EPM2011 to the EPM2017 ephemerides.
 By rescaling $\upsigma_{\dot\varpi}$ by a factor of $\kappa=10$, an uncertainty of $2\times 10^{-6}$ in $\ton{2  + 2\gamma - \beta}/3$ would still be a remarkable result. For the sake of a comparison, in their Table~5, \citet{2014CeMDA.119..237P} claimed $\upsigma_\gamma=6\times 10^{-5},~\upsigma_\beta=3\times 10^{-5}$ obtained with the EPM2011 ephemerides, while \citet{2015CeMDA.123..325F}, who used the INPOP13c ephemerides, quoted $\upsigma_\gamma=7\times 10^{-5},~\upsigma_\beta=5\times 10^{-5}$. More recently, on the basis of the MESSENGER data, \citet{2017AJ....153..121P} yielded $\upsigma_\beta=3.9\times 10^{-5}$,  while \citet{2018NatureG} released $\upsigma_\beta=1.8\times 10^{-5}$. On the other hand, it might suggest that, in fact, a factor $\kappa$ somewhat larger than 10 may be more appropriate; $\kappa = 50$ corresponds to an uncertainty of $1\times 10^{-5}$ in  $\ton{1  + 2\gamma - \beta}/3$. Thus, it should  be somewhat like $10\lesssim \kappa \ll 50$.
\section{Discussion and Conclusions}
Here, I will outline some potential uses of the uncertainties in the planetary orbital rates of change tentatively calculated in Table~\ref{tavola1}.

Since \citet{2018AstL...44..554P}, among other things, modeled also the Solar Lense-Thirring effect assuming its existence as predicted by GTR, the uncertainties of Table~\ref{tavola1}, possibly rescaled by a factor \textcolor{black}{$\kappa \gtrsim 10$}, can be viewed as globally representative of the mismodeling/umodeling in all the standard post-Newtonian dynamics of the solar system including GTR to order $\mathcal{O}\ton{c^{-2}}$, classical N-body effects, oblateness of the Sun, asteroids and TNOs, the uncertainties of the propagation of the electromagnetic waves, and the measurement errors.

Should the gravitomagnetic field of the Sun not be modeled, it would be possible to use the supplementary precessions $\Delta\dot I,~\Delta\dot\Omega,~\Delta\dot\varpi$ of Mercury determined in such a way to try to measure the corresponding Lense-Thirring rates of change to a $\simeq 4\%$ level by disentangling them from the competing classical precessions induced by the Sun's quadrupole mass moment $J_2$. Indeed, both $J_2$ and $\mathbf{S}$ induce long-term precessions on $I,~\Omega$ and $\varpi$ in an arbitrary coordinate system not aligned with the Sun's equator \citep{2011PhRvD..84l4001I}; the expected gravitomagnetic perihelion precession of Mercury amounts to $-2~\textrm{mas~cty}^{-1}$.

The availability of, hopefully, all the extra-precessions $\Delta\dot e,~\Delta\dot I,~\Delta\dot\Omega,~\Delta\dot\varpi$ of as many planets as possible may be useful also in regard to the general relativistic $N-$body effect recently calculated by \citet{2018PhRvL.120s1101W} only for the perihelion which, for Mercury, is of the order of $\simeq 0.1\masy$. Indeed, should it also theoretically affect  the other orbital elements, it would be possible, in principle, to use all the supplementary precessions of more than one planet to separate it from the other lager competing Newtonian and post-Newtonian effects.

Even by rescaling the figures of Table~\ref{tavola1} by a factor of $\kappa = 10$, it would allow one to discard the anomalous perihelion precessions predicted by \citet{2017PhLB..769..281L} on the basis of the emergent gravity theory proposed by \citet{2017ScPP....2...16V} which, for Mercury and Mars, amounts to $0.7\masy$ and $0.09\masy$, respectively. Indeed, \citet{2018AstL...44..554P} did not announce any statistically significant non-zero anomaly in their planetary data reduction. Thus, even if no supplementary perihelion advances are displayed in \citet{2018AstL...44..554P}, it is reasonable to speculate that, should they have been produced, they would have been statistically compatible with zero.

The same conclusion holds also for the anomalous perihelion precession of $\simeq 0.5\masy$ \citep{2003PhRvD..67f4002L}, identical for all of the planets up to terms of order $\mathcal{O}\ton{e^2}$, arising from  the Dvali$-$Gabadadze$-$Porrati (DGP) braneworld model \citep{2000PhLB..485..208D}.

An important use of accurately determined extra-rates $\Delta\dot e,~\Delta\dot I,~\Delta\dot\Omega,~\Delta\dot\varpi$ for, say, Mars and Saturn would consist of much tighter constraints on the location of the putative distant Planet Nine, known also as Telisto, whose gravitational action perturbs all the orbital elements of the known planets. Indeed, such more accurate constraints could be inferred along the lines of what \citet{2017Ap&SS.362...11I} did with just $\Delta\dot\Omega,~\Delta\dot\varpi$ of Saturn determined with the INPOP10a ephemerides by \citet{2011CeMDA.111..363F}. Moreover, such a seemingly only planetological and astronomical topic is, instead, connected also with fundamental physics. Indeed, \citet{2009MNRAS.399..474M} and \citet{2011MNRAS.412.2530B} showed that the pull of a remote, point-like  body located toward the Galactic center is dynamically equivalent to that of the external field effect in the planetary regions of the solar system within the framework of the modified Newtonian dynamics.

Finally, I stress once again the importance that the astronomers responsible for the construction of the planetary ephemerides will determine, hopefully as soon as possible,  accurate supplementary rates of change for all the orbital elements of as many planets along with their uncertainties as possible in view of their wide applications in fundamental physics and beyond. Rates accompanied by their uncertainties would be useful for testing various ideas in gravitational physics. Uncertainties alone might be used for planning purposes and sensitivity analyses, but not for tests.
\section*{Acknowledgements}
I am grateful to the anonymous referee for the competent and useful critical remarks.
\bibliography{Gclockbib,semimabib,PXbib}{}
\end{document}